\documentstyle[prl,aps,epsfig,twocolumn]{revtex}
\begin{document}
\draft

\title{Laser cooling all the way down to Fermi superfluid}

\author{J. Dziarmaga$^1$ and  M. Lewenstein$^2$}

\address{
$^1$ 
Intytut Fizyki Uniwersytetu Jagiello\'nskiego, 
ul.~Reymonta 4, 30-059 Krak\'ow, Poland \\
$^2$
Institut f\"ur Theoretische Physik,
Universit\"at Hannover, 
D-30167 Hannover, Germany
}

\date{ October 26, 2004 }
\maketitle

\begin{abstract}
Numerical simulations for realistic experimental parameters demonstrate 
that laser cooling on the attractive side of the Feshbach resonance can 
drive fermions much below the superfluid transition. For the assumed 
set of experimental parameters the transition takes place at 
$0.35T_F$, and laser cooling can drive the system down to at least 
$0.085T_F$ in a time of a few seconds. Superfluid growth is 
self-consistently included in simulations. 
\end{abstract}


Fermi superfluids are at the forefront of both experimental and
theoretical activity in the physics of ultracold dilute atomic gases. In
addition to being a challenge in their own right, they provide promising
model systems to test the theory of fermionic superfluidity, or the
Bardeen-Cooper-Schrieffer (BCS) theory, in the regime of strong
interactions, when the size of Cooper pairs becomes comparable, or even less
than the average interparticle distance. In this regime excitations of the
system develop a pseudogap indicating that Cooper pairs are quasi bound
states of pairs of fermions \cite{Randeria}. This is also the limit that
seems to be a key to understanding the physics of high $T_c$
superconductors, where pseudogap effects have been observed experimentally.  
The cold atoms provide a model system of high flexibility thanks to
application of the Feshbach resonance technique \cite{Feshbach}. The
Feshbach resonance between an open and closed channel allows one to
manipulate an effective s-wave scattering length between pairs of
atoms with opposite ``spin''.  This technique allows one to drive the
system through the BCS-BEC crossover, when the scattering length diverges
and changes sign from negative (attractive) to positive (repulsive). At
the same time the state of the system continuously evolves from the BCS
superfluid of weakly correlated Cooper pairs, through the pseudogap
regime, and then the unitarity limit, when the scattering length diverges, 
to the Bose-Einstein condesate of biatomic
molecules.

The main problem in the experiments on atomic Fermi superfluids is how to
cool the system sufficiently deep into the quantum degenerate regime to
reach the superfluidity. Unfortunately the standard evaporative cooling
stops to work in the degenerate regime, but before the superfluidity is
achieved \cite{OHara}. One of the reasons may be creation of holes deep in
the Fermi sea by collisions with the buffer gas \cite{heating}. However,
this problem can be circumvented in an ingenious way thanks, again, to the
Feshbach technique \cite{ingenious}. The idea is to obtain first a
molecular condensate on the BEC side of the Feshbach resonance, and then
continuously transform it into BCS superfluid by an adiabatic passage to
the other side of the resonance. This adiabatic transition reduces the
temperature sufficiently to reach the superfluid state. By now, several
succesful experiments with molecular condesates were performed,
and paved a way to first succesful experiments with Fermi superfluids
\cite{mBEC,Grimm}. The Innsbruck experiment \cite{Grimm} is the first to report
direct signatures of Fermi superfluidity in
quasiparticle excitation spectrum, in excellent agreement with the
theory of Refs.\cite{Torma}.

The purpose of this Letter is twofold. First, we
 demonstrate that it is possible to laser
cool a Fermi gas down to the superfluid BCS state on the 
attractive (BCS) side of the Feshbach resonance.  This 
is challenging because, so far, the standard laser cooling has
not even allowed for observing quantum degeneracy. However, theoretical
studies of laser cooling show in particular that it is possible to
reach the degeneracy in Fermi systems \cite{Idziaszek_f}. Laser cooling
offers an alternative route towards BCS superfluid directly on the
negative side of the Feshbach resonance. Moreover, it opens a
possibility to cool the BCS superfluid prepared via the ``standard''
method down to even lower temperatures. The second goal of this Letter 
is to develop for
the first time fully self-consistent quantum kinetic theory of BCS state
creation via laser cooling \footnote{similar to self-consitent theories of
the BEC growth \cite{GZ}}. The results are very promising: despite the
fact that both BCS formation and normal mean field effects shift the
energies of quasiparticle excitations with respect to the equally spaced
harmonic oscillator levels, the cooling remains efficient at least down to
the temperature  $T\simeq 0.085T_F$. This temperature is well below the
superfluid transition temperature which for the assumed reasonable
experimental parameters is close to $0.35T_F$. Optimization of the cooling
protocol could  significantly reduce this temperature.


We consider two species of fermions with ``spin up and down'', molecules
which are their bound states, and fermions in an excited state.
Populations of the spin up and down fermions are the same. We assume that
mutual s-wave interaction between these two populations leads to
thermalisation on a timescale, which is faster than the rate of laser
cooling. The system remains in a quasi-equilibrium state with slowly
decreasing temperature. This equlibrium state is described by the
generalised weak coupling BCS theory, or boson-fermion model \cite{BF}
which includes Bose-Einstein condensate of the molecules. Structure of the
ground state and its excitations follows from the set of Bogoliubov-de
Gennes equations together with self-consistent definitions of the gap
function and the mean field potential. Laser cooling of fermions was
described in detail in Ref.\cite{Idziaszek_f}. Coherent laser excites
atoms from, say, the spin down ground state to the excited state and at
the same time spontaneous emission brings them back to the spin down
ground state. Frequencies of the laser assure that the generalized Raman
cooling takes place \cite{Idziaszek_b}. In this Letter we take into
account interactions between fermions and describe the cooling process in
terms of instantaneous Bogoliubov quasiparticles, whose eigenfunctions and
eigenenergies are self-consistently updated during the evolution. We work
in the {\it festina lente} limit to avoid reabsorbtion effects \cite{FL},
and also employ spherical symmetry and ergodic approximations(see also 
\cite{GZ}).

The boson-fermion model \cite{BF} is defined by the Hamiltonian
\begin{eqnarray}
\hat H=
&\int d^3r&
\left[
\sum_{a=+,-,e}
\hat\psi^{\dagger}_a
{\cal H}_1
\hat\psi_a+
g_0
\hat\psi_{+}^{\dagger}
\hat\psi_{-}^{\dagger} 
\hat\psi_{-}
\hat\psi_{+}+
\right.
\nonumber\\
&&
\left.
\lambda
\hat\psi^{\dagger}_m
\hat\psi_{+}
\hat\psi_{-}+   
{\rm ~h.c.}+
(2\nu-2\mu)   
\hat\psi^{\dagger}_m\hat\psi_m
\right]~.
\label{HBF}  
\end{eqnarray}
The fields $\hat\psi_{+}$ and $\hat\psi_{-}$ describe the fermions with
spin up and down respectively, $\hat\psi_e$ corresponds to the fermions
in the excited state, and $\hat\psi_m$ is the bosonic molecular field.
Here ${\cal H}_1=-\frac{\hbar^2}{2m}\nabla^2+V(\vec r)-\mu$ is a
single particle Hamiltonian, $m$ is atomic mass, $V(\vec r)$ is a
harmonic trap potential, $\mu$ is chemical potential,
$g_0=\frac{4\pi\hbar^2 a_0}{m}$ is bare interaction strength with a
negative bare $s$-wave scattering length $a_0$, $\lambda$ is a coupling 
between pairs of atoms and molecules, and $\nu$ is detuning from the 
Feshbach resonance.

In the mean field approximation, which closely follows the BCS
theory, the quartic interaction term in the Hamiltonian (\ref{HBF}) is
made quadratic by replacing all products of pairs of 
operators with their averages. The averages are the mean field potential
$W(\vec r)=
g_0
\left\langle
\hat\psi^{\dagger}_{\pm}(\vec r)
\hat\psi_{\pm}(\vec r)
\right\rangle $,
and the anomalous potential of the pairing field 
$P(\vec r)=
\left\langle
\hat\psi_+(\vec r)
\hat\psi_-(\vec r)
\right\rangle $ ,
which is mixing fermions with spin up and down. The molecular field
$\hat\phi_m$ is replaced by the amplitude of molecular condensate
$\langle\hat\phi_m\rangle$. After ellimination of the molecular
condensate one obtains linear equations for the spin up and
down fermions
\begin{eqnarray}
&&
i\hbar\frac{d}{dt}~\hat\psi_{\pm}~=~
{\cal H}_1~\hat\psi_{\pm}+
W(\vec r)~\hat\psi_{\pm}~\mp~
g~P(\vec r)~
\hat\psi^{\dagger}_{\mp}~,
\label{pm2}
\end{eqnarray}
with an effective coupling constant
$
g~=~g_0~+~\frac{\lambda^2}{2(\mu-\nu)}~.
$
These equations mix $\hat\psi$ and $\hat\psi^{\dagger}$, but
can be ``diagonalized'' by the Bogoliubov transformation
\begin{eqnarray}
&&
\hat\psi_+(\vec r)~=~
\sum_m~
\hat b_{m,+}~           u_m(\vec r) ~-~
\hat b_{m,-}^{\dagger}~ v_m^*(\vec r) ~,
\label{Bog+}\\
&&
\hat\psi_-(\vec r)~=~
\sum_m~
\hat b_{m,-}~           u_m(\vec r) ~+~
\hat b_{m,+}^{\dagger}~ v_m^*(\vec r) ~,
\label{Bog-}
\end{eqnarray}
with fermionic quasiparticle annihilation operators $\hat b_{m,\pm}$.
The Bogoliubov modes $(u_m,v_m)$ fulfill the Bogoliubov-de Gennes
equations
\begin{eqnarray}
&&
\omega_m~u_m~=~ 
+~{\cal H}_1~u_m~+~
W~u_m~
+~\Delta~v_m~,
\nonumber\\
&&
\omega_m~v_m~=~ 
-~{\cal H}_0~v_m~-
~W~v_m~
+~\Delta^*~u_m ~,
\label{BdG}
\end{eqnarray}
with positive energies $\omega_m$ of quasiparticle excitations. Here
$\Delta=-gP$ is the quasiparticle gap function. In a thermal state with
inverse temperature $\beta$, the average occupation
numbers of quasiparticle states are given by Fermi-Dirac distribution
$N_m=\langle \hat b^{\dagger}_{m,\pm} \hat b_{m,\pm} \rangle=
(1+\exp(\beta\omega_m ))^{-1}$. Equations (\ref{BdG}) are solved
together with the self-consistency conditions
\begin{eqnarray}
\Delta(\vec r)&=&
\left[
-g
\sum_m
\left( 1 - 2 N_m \right) u_m(\vec r) v_m^*(\vec r) 
\right]_{\rm reg}~,  
\label{Puv}\\
W(\vec r)&=&
g_0
\sum_m
(1-N_m) |v_m(\vec r)|^2 +
N_m     |u_m(\vec r)|^2~,
\label{Wuv}
\end{eqnarray}
by succesive iterations with the chemical potential $\mu$ adjusted to keep
the average number of atoms constant. The ultraviolet divergence in
Eq.(\ref{Puv}) is regularized using the quickly convergent method of
Ref.\cite{LDA}.

Excitation of atoms from the spin down state to the excited state is
described by the Hamiltonian
\begin{equation}
\hat H_{\rm las}=
\int d^3r~
\left[
\frac12\Omega
e^{i\vec k_{L}\vec r}
\hat\psi_{e}^{\dagger}(\vec r)
\hat\psi_{-}(\vec r)+
{\rm h.c.}-
\delta
\hat\psi^{\dagger}_e
\hat\psi_e
\right],
\end{equation}
driving coherent oscillations with Rabi frequency $\Omega$ and laser
detuning $\delta$. The excitation is accompanied by the spontaneous
emission, which after rotating wave approximation is described by a 
superoperator
\begin{eqnarray}
{\cal L}~\hat\rho&=&
\gamma 
\sum_{mk} 
U_{mk}
{\cal D}[\hat b_{m,-}^{\dagger}\hat e_k]\rho +
V_{mk}
{\cal D}[\hat b_{m,+}\hat e_k]\rho,
\end{eqnarray}
with a spontaneous emission rate of $\gamma$. Here the Lindblad
superoperator is
${\cal D}[\hat A]\rho=\hat A\rho\hat A^{\dagger}-
\frac12\rho\hat A^{\dagger}\hat A-\frac12\hat A^{\dagger}\hat A\rho$, and
the matrix elements are e.g.
$
U_{mk}=
\int d\Omega_k  {\cal W}(\Omega_k)
|u_{mk}(\vec k)|^2
$
with the spontaneous emission pattern ${\cal W}(\Omega_k)$ and the
generalized Frank-Condon factors
$
u_{mk}(\vec k)=
\int d^3r~
e^{i\vec k \vec r}
w_k^*(\vec r)
u_m(\vec r)
$.
Here $w_k(\vec r)$ is the $k$-th eigenstate of the harmonic 
oscillator and $\hat e_k$ an annihilation operator of an excited
atom in this state. 

Adiabatic ellimination of the excited state, similar as in
Ref.\cite{AE}, leads to the kinetic equations for the occupation
numbers taking into account the effects of Fermi statistics
\begin{eqnarray}
\frac{d N_{m,-}}{dt} &=&
\sum_n~
\Gamma^{(-)}_{m\leftarrow n} 
\left( 1 - N_{m,-} \right) N_{n,-} -
(n\leftrightarrow m)+
\nonumber\\
&&
C_{mn} \left( 1 - N_{m,-} \right) \left( 1 - N_{n,+} \right) -
A_{nm} N_{m,-} N_{n,+} ~,
\nonumber\\
\frac{d N_{m,+}}{dt} &=&
\sum_n~
\Gamma^{(+)}_{m\leftarrow n} 
\left( 1 - N_{m,+} \right) N_{n,+} -
(n\leftrightarrow m)+
\nonumber\\
&&
C_{nm} \left( 1 - N_{m,+} \right) \left( 1 - N_{n,-} \right) -
A_{mn} N_{m,+} N_{n,-} ~.
\nonumber
\end{eqnarray}
They describe relaxation of $\pm$ quasiparticles
$(\Gamma^{(\pm)}_{m\leftarrow n})$, and creation/annihilation of pairs of
$+$ and $-$ quasiparticles $(C_{nm}{\rm ~and~ }A_{nm})$. The transition 
rates are e.g.
\begin{equation}
\Gamma^{(-)}_{m\leftarrow n}=
\frac{\Omega^2}{2\gamma}
\sum_k
\frac{
\gamma^2 
U_{mk} 
|u_{nk}(\vec k_L)|^2
}
{
(\delta-\omega^e_k+\omega_n)^2+
\gamma_k^2
}~.
\label{Gamma-}
\end{equation}
Here $\omega^e_k$ is the energy of the $k$-th harmonic oscillator
state and $\gamma_k$ is approximate spontaneous decay rate of an
excited atom in this state:
$\gamma_k=\gamma\sum_mU_{mk}(1-N_{m,-})+V_{mk}N_{m,+}$.

Laser cooling drives average occupation numbers $N_{m,\pm}$ of
quasiparticle states out of thermal equilibrium. At the same time
interactions between quasiparticle states drive the system towards
thermal equilibrium. Numerical simulations in Ref.\cite{Idziaszek_f}
show that for cooling rates in ms range thermal relaxation remains 
faster than cooling all the way down to $T\simeq 0.03~T_{\rm F}$. 
This justifies our assumption of
fast equilibration to a quasi-equilibrium state. After each short
period of laser cooling the occupation numbers $N_{m,\pm}$ go out of
equilibrium, where $N_{m}=[1+\exp(\beta\omega_m)]^{-1}$. The initial total
energy $E(T)~=~\sum_{m,\pm}\omega_m N_{m,\pm}$ changes by
$dE~=~\sum_{m} \omega_m~(N_{m,+}+N_{m,-}-2N_m)$. The relaxation
brings the occupation numbers $N_{m,\pm}$ to a new state of
equilibrium at a temperature $T+dT$, but it does not change the total
energy. The energy of the system at the new temperature $E(T+dT)$ differs
only slightly from the initial $E(T)$ by $dE_{T}=E(T+dT)-E(T)\approx
\frac{2dT}{T^2}\sum_m\omega_m^2 N_m(1-N_m)$. Conservation of energy
in thermal relaxation means that $dE=dE_{T}$. In our simulations we use
this equality to find the new lower temperature $T+dT$, and then solve
the Bogoliubov-de Gennes equations self-consistently to adjust the
Bogoliubov modes and the chemical potential to the lower temperature.
With the new Bogoliubov modes we calculate new transition rates etc.

Calculation of the transition rates is the most time consuming part of the
numerical simulation. This is why we were forced to assume spherical
symmetry. For spherically symmetric $\Delta(r)$ and $W(r)$ the Bogoliubov
modes can be separated as e.g. $u_{ln}(r)~Y_{lm}(\theta,\varphi)$ and
their energies $\omega_{ln}$ do not depend on $m$. To be consistent with
the spherical symmetry we also made the ergodic 
approximation \cite{Idziaszek_f,GZ}, assuming that 
quasiparticle occupation numbers $N_{ln,\pm}$ do not depent on $m$. In
other words, we assume fast equilibration within each quasiparticle energy
shell, which keeps the system in a spherically symmetric state.

In our simulations we use 81 harmonic oscillator levels (shells) and assume
that there are $N=10660$ atoms with spin up and down. In the
non-interacting case at $T=0$ the atoms  fill 39  energy
shells. The frequency of the isotropic trap is $\omega=2\pi~2400$Hz.
The bare scattering length for the interaction between the two species is
$a_0=-157a_B$, where $a_B$ is the Bohr radius. This value
corresponds to the interactions between the spin states
$|F=9/2,m_F=9/2\rangle$ and $|F=9/2,m_F=7/2\rangle$ of $^{40}$K. In
the harmonic oscillator units this scattering length gives
$g_0=-0.32$. In our calculations we use
two values for the effective interaction strength: $g=-0.32$ and
$g=-1$ enhanced by the Feshbach resonance. The wavelength of the
cooling laser is $\lambda=720$nm and the laser detuning is
$\delta=-12\hbar\omega$. In all simulations the Rabi frequency 
is much less than the spontaneous emission rate,
$\Omega=0.1\gamma$, so that average occupation of excited
state remains small. This is one of the assumptions in the adiabatic
ellimination of the excited state.

\begin{figure}
\centering  
{\epsfig{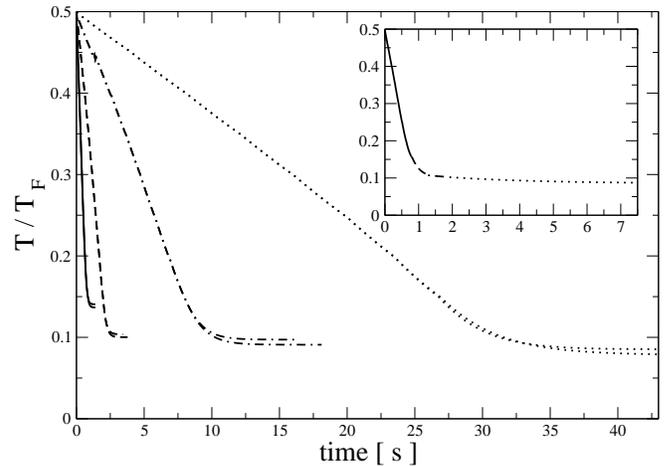}}
\caption{
Temperature as a function of time during the laser cooling for four
values of $\gamma$: $\gamma=10$ (solid lines), $\gamma=5$ (dashed lines), 
$\gamma=2.5$ (dashed-dotted lines), and $\gamma=1.25$ (dotted lines). Each 
pair of plots in this figure corresponds to $g=-0.32$ (higher temperature) 
and $g=-1$ (lower temperature). The lowest temperatures of $0.079T_F$ and 
$0.085T_F$ respectively can be reached for $\gamma=1.25$. On the other 
hand, the initial stage of laser cooling is efficient at most for 
$\gamma=10$, when the temperature of $0.15T_F$ is reached within $0.85$s. The 
inset shows laser cooling to the superfluid state with $g=-1$
realized in four stages to reach the lowest temperature in reasonable
time. The spontaneous emission rate is switched as $\gamma=10\to 5\to
2.5 \to 1.25$ at the temperatures $T/T_F=0.16,0.11,0.10$, correspondingly.}
\label{T_t}
\end{figure}

\begin{figure}
\centering
\vspace*{1cm}
{\epsfig{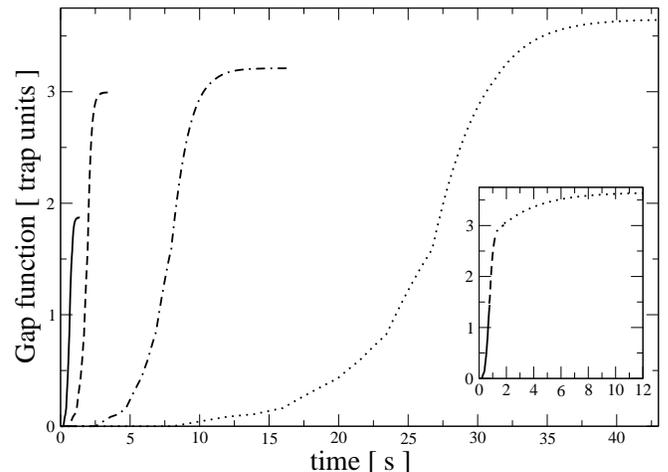}}
\caption{
The maximal magnitude of the gap function in the center of the trap
$|\Delta(r=0)|/\hbar\omega$ for the same simulations with $g=-1$ as in 
Fig. \ref{T_t}. The inset shows the gap in the four stage cooling of 
the inset of Fig.\ref{T_t}.
}
\label{Delta_t}
\end{figure}

In Fig.\ref{T_t} we show temperature as a function of time for laser
cooling with four values of $\gamma$ consistent with the {\it festina
lente} regime. Each plot in this figure consists in fact of 
two indistinguishable
plots, one for the bare $g=-0.32$ and the other for the enhanced $g=-1$.  
For the stronger $g=-1$ there is a transition to the superfluid state at
low temperature (see Fig.\ref{Delta_t}), but the non-zero gap function
which appears at the transition does not influence the efficiency of laser
cooling. The process of laser cooling is dominated by the higly degenerate
states with large angular momenta $l\approx E_F/\hbar\omega$ whose
quasiparticle energies are not influenced much by the gap function
localized in the center of the trap \cite{Torma,Baranov}. In more physical
terms, the superfluid droplet in the center of the trap is surrounded by a
shell of normal Fermi gas. As long as this mesoscopic effect persists the
laser cooling in the superfluid state remains as efficient as in the
normal state.

The initial cooling is the fastest for $\gamma=10$ but the lowest
temperature of $0.085T_F$ is obtained for $\gamma=1.25$. The best
strategy to reach low temperatures in reasonable time is to do the
cooling in a few stages starting with large $\gamma$ and finishing
the job with small $\gamma$, or to continuously decrease $\gamma$ as
the temperature decreases. The inset in Fig.\ref{T_t} shows an
example of a four stage cooling process.

Summarizing, we made numerical simulations for reasonable experimental
parameters consistent with the requirements of the {\it festina lente}
regime which demonstrate that it is possible to laser cool fermions on the
negative side of the Feshbach resonance all the way down to the Fermi
superfluid. Despite the pairing effects and the usual mean field
interaction, which shift quasiparticle energies with respect to the equally
spaced harmonic oscillator levels, it is experimentally fisible to reach at
least $0.085T_F$ in a time as short as a few seconds.


We would like to thank Misha Baranov,
Luis Santos, and Zbyszek Idziaszek for discussions. This work was
supported in part by ESF QUDEDIS programme, DFG (SFB 407), and the KBN
grant PBZ-MIN-008/P03/2003.



\end{document}